\documentclass{PoS}
\let\ifpdf\relax
\usepackage{amsmath}
\usepackage{graphicx}
\usepackage{bm}

\title{Static quark-antiquark pair free energy and screening masses: continuum results at the\\ QCD physical point}

\ShortTitle{Static $Q\bar{Q}$ pair free energy and screening masses: continuum results at the physical point}

\author{Szabolcs Bors\'anyi\\
        University of Wuppertal }

\author{Zolt\'an Fodor\\
        University of Wuppertal }

\author{S\'andor D. Katz\\
E\"otv\"os University and MTA-ELTE Lend\"ulet Lattice Gauge Theory Research Group}

\author{\speaker{Attila P\'asztor}\\
       University of Wuppertal\\
        E-mail: \email{apasztor@bodri.elte.hu}}

\author{K\'alm\'an K. Szab\'o\\
University of Wuppertal and J\"ulich Supercomputing Center}

\author{Csaba T\"or\"ok\\
E\"otv\"os University and MTA-ELTE Lend\"ulet Lattice Gauge Theory Research Group}

\abstract{ We study the correlators of Polyakov loops, and the corresponding gauge 
invariant free energy of a static quark-antiquark pair in 2+1 flavor QCD at finite 
temperature. Our simulations were carried out on $N_t$ = 6, 8, 10, 12, 16 lattices using a 
Symanzik improved gauge action and a stout improved staggered action with physical quark 
masses. The free energies calculated from the Polyakov loop correlators are extrapolated 
to the continuum limit. For the free energies we use a two step renormalization procedure 
that only uses data at finite temperature. We also measure correlators with definite 
Euclidean time reversal and charge conjugation symmetry to extract two different screening 
masses, one in the magnetic, and one in the electric sector, to distinguish two 
different correlation lengths in the full Polyakov loop correlator. This conference 
contribution is based on the paper: JHEP 1504 (2015) 138
}

\FullConference{The 33rd International Symposium on Lattice Field Theory\\
                 14 -18 July  2015\\
                 Kobe International Conference Center, Kobe, Japan}

\begin{document}

\section{Introduction}
At high temperatures strongly interacting matter undergoes a
transition where colorless hadrons turn into a phase dominated by
colored quarks and gluons, the quark gluon plasma (QGP).
Deconfinement properties of the transition can be studied by
infinitely heavy, static test charges. Here, we calculate
the gauge invariant static quark-antiquark pair free energy.

The free energy of a static quark-antiquark pair as a function of their distance at various temperatures
is determined by the Polyakov loop correlator~\cite{McLerran:1981pb}, which gives the 
gauge invariant $\bar{Q}Q$ free energy \footnote{More precisely, the excess free energy that 
we get when inserting two static test charges in the medium.} as:
\begin{equation} \label{def_of_F}
F_{\bar{Q}Q}(r) = -T \ln C(r,T) = - T \ln \left\langle \sum_{\textbf{x}} \mathrm{Tr} L(\textbf{x}) \mathrm{Tr} L^+(\textbf{x}+\textbf{r}) \right\rangle \rm{.}
\end{equation}
In the above formula, $\textbf{x}$ runs over
all the lattice spatial sites, and the Polyakov loop, $L(\textbf{x})$,
is defined as the product of temporal link variables\footnote{In the literature, a factor of
$\frac{1}{N_c}$ is often included in the definition. Including this factor leads to a term in the static quark free
energy that is linear in temperature.} $U_4(\textbf{x},x_4) \in SU(3)$:
\begin{equation}
L(\textbf{x})=\prod_{x_4=0}^{N_t-1} U_4 (\textbf{x},x_4) \rm{,}
\end{equation} 

A related problem is distinguishing correlation lengths in the correlator
of Polyakov loops, which give inverse screening masses in the plasma. In the
full Polyakov loop correlator the electric and magnetic sectors both 
contribute.
In order to investigate the effect of electric and magnetic gluons separately, 
one can use the symmetry of Euclidean time reflection~\cite{Arnold:1995bh}, 
that we will call ${\cal R}$. The crucial property of magnetic versus
electric gluon fields $A_4$ and $A_i$ is that under this symmetry, one is intrinsically 
odd, while the other is even:
\begin{equation}
 A_4(\tau,\textbf{x}) \xrightarrow{{\cal R}} - A_4(-\tau,\textbf{x})\rm{,} \quad A_i(\tau,\textbf{x}) \xrightarrow{{\cal R}} A_i(-\tau,\textbf{x})
\end{equation}
Under this symmetry the Polyakov loop transforms as $L \xrightarrow{{\cal R}} L^{\dagger}$. One can easily define correlators that are even or odd under 
this symmetry, and thus receive contributions only from the magnetic or electric sector, 
respectively~\cite{Arnold:1995bh, Maezawa:2010vj}: \\
\begin{eqnarray}
 L_M \equiv (L + L^\dagger)/2 \quad \quad L_E \equiv (L - L^\dagger)/2 \rm{.}
\end{eqnarray}
We can further decompose the Polyakov loop into ${\cal C}$ even and odd states, using 
$A_4 \xrightarrow{{\cal C}} A_4^*$ and $L \xrightarrow{{\cal C}} L^*$ as:
\begin{eqnarray}
L_{M\pm} = (L_M \pm L^*_M)/2 \quad \quad L_{E\pm} = (L_E \pm L^*_E)/2 \rm{.}
\end{eqnarray}
Next, we note that $\operatorname{Tr} {L_{E+}}=0=\operatorname{Tr} {L_{M-}}$, so the decomposition of the Polyakov loop correlator to 
definite ${\cal R}$ and ${\cal C}$ symmetric operators contains two parts\footnote{Note that the Polyakov loop correlator does 
not overlap with the ${\cal R}({\cal C})=+(-)$ and ${\cal R}({\cal C})=-(+)$ sectors. To access these sectors, other
operators are needed.}. We define the magnetic correlation function as:
\begin{equation}
\label{eq:Cmag}
C_{M+}(r,T) \equiv \left\langle \sum_{\bf x} \mathrm{Tr} L_{M+}({\bf x}) \mathrm{Tr} L_{M+}({\bf x}+{\bf r}) \right\rangle - \left| \left\langle \sum_{\bf x} \mathrm{Tr} L({\bf x}) \right\rangle \right|^2 \rm{,}
\end{equation} 
\\
and the electric correlator as\footnote{Here our definition differs from that used in~\cite{Maezawa:2010vj} in a sign.}:
\\
\begin{equation}
\label{eq:Cel}
C_{E-}(r,T) \equiv - \left\langle \sum_{\bf x} \mathrm{Tr} L_{E-}({\bf x}) \mathrm{Tr} L_{E-}({\bf x}+{\bf r}) \right\rangle \rm{.}
\end{equation}

Then, from the exponential decay of these correlators, we can define the magnetic 
and electric screening masses. Note that with our definition $\mathrm{Tr}L_{M+} = \operatorname{Re} \mathrm{Tr}L$  and 
$\mathrm{Tr}L_{E-} = i \operatorname{Im} \mathrm{Tr}L$ , and:
\begin{equation}
\label{eq:corr_eq_mag_plus_el}
C(r,T) - C(r \to \infty,T) = C_{M+}(r,T) + C_{E-}(r,T) \rm{,}
\end{equation}
from which it trivially follows that if the magnetic mass screening mass is lower than the electric mass, we will 
have $C(r,T) - C(r \to \infty,T)$ asymptotic to $C_{M+}(r,T)$ as $r \to \infty$, or equivalently, the highest 
correlation length in $C$ equal to that of $C_{M+}$. We will determine the correlation lengths by fitting a
Yukawa ansatz to these correlators.

\section{Simulation details}

The simulations were performed by using the tree level Symanzik improved gauge, 
and stout-improved staggered fermion action, that was used in~\cite{Aoki:2005vt}.
We worked with physical quark masses, and fixed them by reproducing the 
physical ratios $m_{\pi}/f_K$ and $m_K/f_K$.

Compared to our previous investigations of Polyakov loop correlators, reported 
in the conference proceedings ~\cite{Fodor:2007mi}, here we used finer lattices, 
namely we carried out simulations on $N_t=12$ and $16$ lattices as well as 
on $N_t=6, 8, 10$ lattices. Our results were obtained in the temperature 
range 150 MeV $\leq$ T $\leq$ 450 MeV. We use the same configurations as in 
\cite{Borsanyi:2010bp} and \cite{Borsanyi:2013bia}.

\section{The gauge invariant free energy}

\subsection{Renormalization}

We use a renormalization procedure based entirely on our $T>0$ data, similarly to 
Refs. ~\cite{Gupta:2007ax} and ~\cite{Borsanyi:2012uq}. The data contains a temperature 
independent divergent part from the ground state energy. The difference between the 
value of free energies at different temperatures is free of divergences.
Accordingly, we define the renormalized free energy as:

\begin{equation} \label{eq:renorm}
F_{\bar{Q}Q}^{ren} (r,\beta,T;T_0) = F_{\bar{Q}Q} (r,\beta,T) - F_{\bar{Q}Q}(r \rightarrow \infty, \beta, T_0) \rm{,}
\end{equation} 
\\
with a fixed $T_0$. This renormalization prescription corresponds to the choice that
the free energy at large distances goes to zero at $T_0$. We choose $T_0=200$MeV. 
Our renormalization procedure is implemented in two steps.

In the first step we renormalize the single static quark free energy which satisfies:
\begin{equation}
\label{eq:F1Q}
2 F_{Q} (\beta, T) = F_{\bar{Q}Q} (r \to \infty,\beta,T) = -T \log \left| \left< \operatorname{Tr} L  \right>\right|^2 \rm{.}
\end{equation}
We define its renormalized counterpart as:
\begin{equation}
\label{eq:renorm1Q}
F^{ren}_Q(\beta,T; T_0) = F_{Q}(\beta,T)-F_{Q}(\beta,T_0) \rm{.}
\end{equation}
In the second step the full renormalized $\bar{Q}Q$ free energy can be written as:
\begin{equation}
F_{\bar{Q}Q}^{ren}(r,\beta,T;T_0) = \tilde{F}_{\bar{Q}Q}(r,\beta,T) + 2 F^{ren}_Q (\beta, T; T_0) \rm{,}
\end{equation}
where
\begin{equation}
\tilde{F}_{\bar{Q}Q}(r,\beta,T) = F_{\bar{Q}Q} (r,\beta,T) - F_{\bar{Q}Q} (r \to \infty,\beta,T)  = F_{\bar{Q}Q} (r,\beta,T) - 2 F_{Q} (\beta,T)  \rm{.}
\end{equation}
Note, that this second step of the renormalization procedure is completely straightfoward to implement, at
each simulation point in $N_t$ and $\beta$ we just subtract the asymptotic value of the correlator.
The main advatage of this 2 step procedure is that it allows us to extend the temperature range 
we can do a continuum limit in without performing $T=0$ simulations at lots of different $\beta$
values. For more details see \cite{Borsanyi:2015yka}.

\subsection{Smearing as variance reduction}

\begin{figure}[t!]
\begin{center}
\includegraphics[width=0.7\textwidth, angle=0]{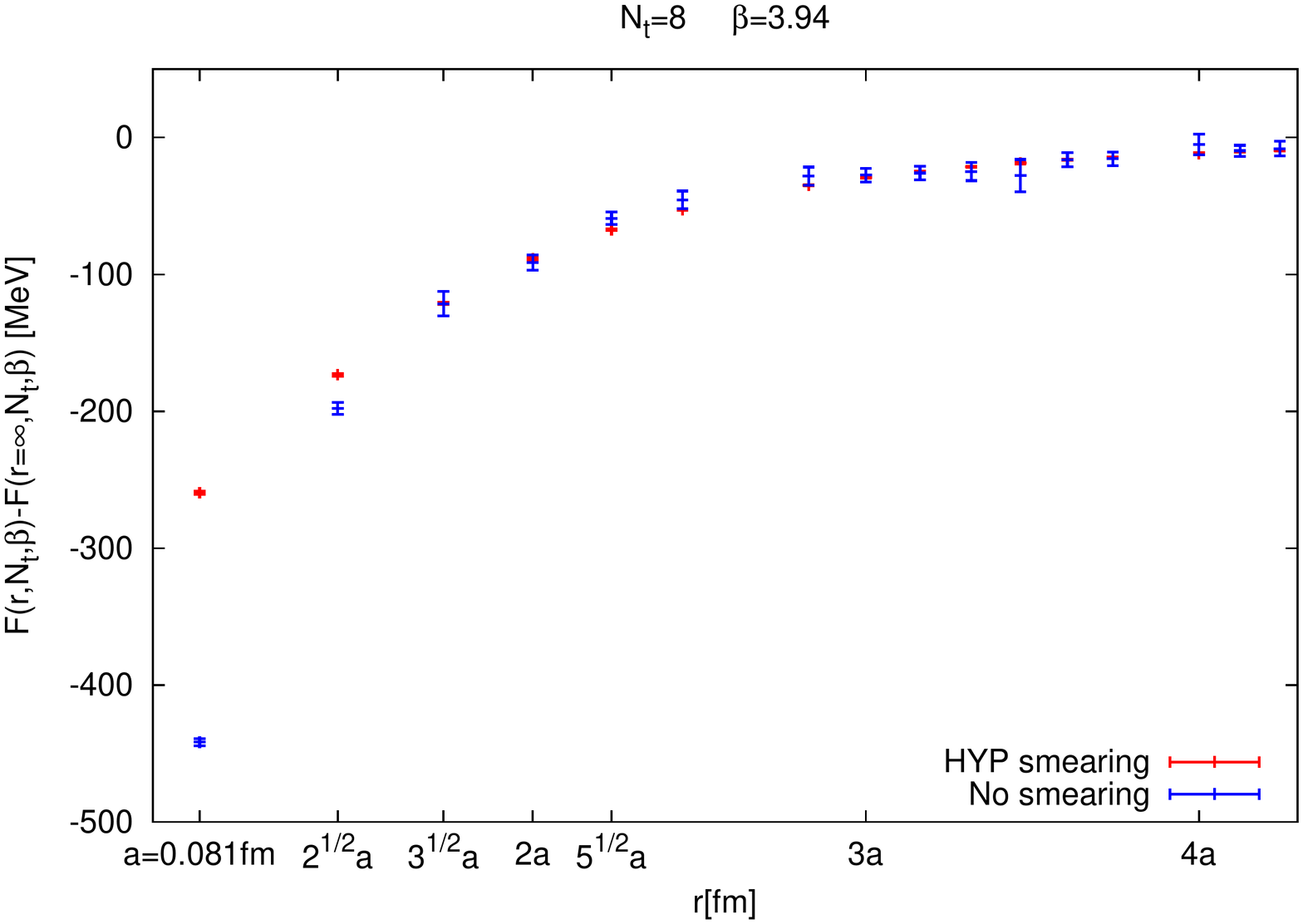}
\end{center}
\vspace{-1.4cm}
\caption{The smeared and unsmeared free energies at a given $\beta$ and $N_t$, after the first step of the renormalization procedure. }
\label{fig:smear}
\end{figure}

The Polyakov loop correlator behaves similarly to baryon correlators in imaginary time: at large values 
of $r$ we can get negative values of $C$ at some configurations\footnote{Of course, the ensemble average should in 
principle be positive definite.}. For this reason, it is highly desirable to use gauge field smearing which makes 
for a much better behavior at large $r$, at the expense of unphysical behavior at small $r$. For this reason, 
we measured the correlators both without and with HYP smearing. We expect that outside the smearing range 
(i.e. $r \geq 2a$) the two correlators coincide. This is supported by Figure \ref{fig:smear}. Therefore we use the 
smeared correlators for $r \geq 2a$ and the unsmeared ones for $r<2a$.

\subsection{Results}

\begin{figure}[t!]
\begin{center}
\includegraphics[width=0.7\textwidth, angle=0]{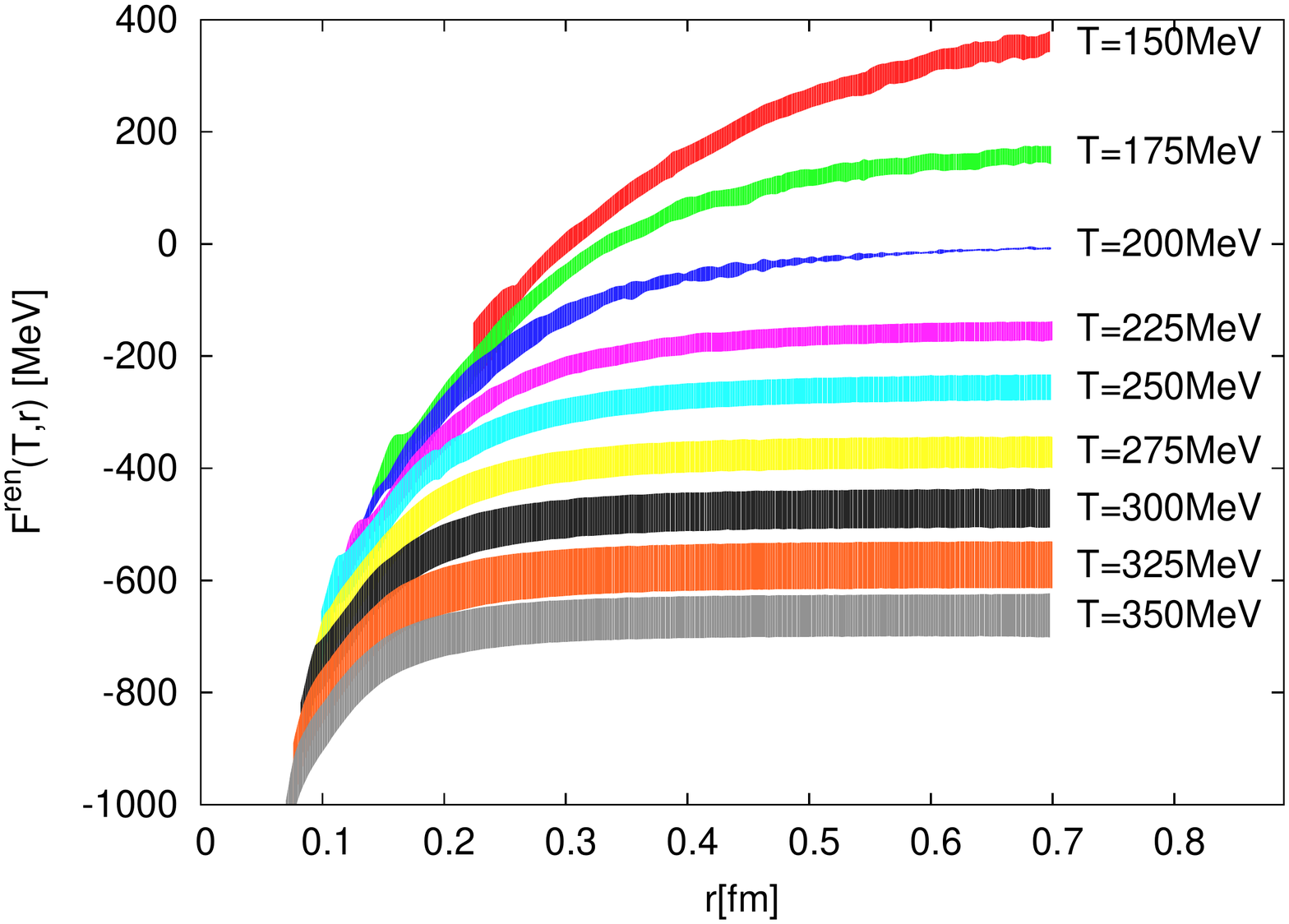}
\vspace{-1cm}
\caption{Continuum values of the static $\bar{Q}Q$ free energy at different temperatures. }
\vspace{-1cm}

\label{fig:final_free_energy}
\end{center}
\end{figure}

The continuum extrapolations were done with $N_t=8,10,12$ and when available $N_t=16$ lattices.
For details on the systematic error estimation see \cite{Borsanyi:2015yka}.
The final results are in Figure \ref{fig:final_free_energy}. Note that the curves seem to tend to
the same curve as $r \to 0$, corresponding to the expectation that UV physics is temperature independent.
Also note that the error bars get smaller as we approach $T_0=200MeV$, which was chosen as a 
renormalization point. This is a natural consequence of the implementation of our renormalization
prescription. It is also the reason why the correlator tends to zero at that point.
A different renormalization would correspond to a constant shift in this graph.

We note, that on this conference an other determination of the same free energy was reported \cite{SameConf}
using the HISQ action, and there is a slight difference in the asymptotic value (or the single quark free energy), 
at the higher temperature values of $T>300$MeV. It amounts to approximately $1-1.5\sigma$. In the lower
temperature range, where published continuum data on the Polyakov loop is also available, we have an 
agreement, see e.g. \cite{Bazavov:2013yv}.

\section{The screening masses}

\begin{figure}
\includegraphics[angle=0, width=0.48\linewidth]{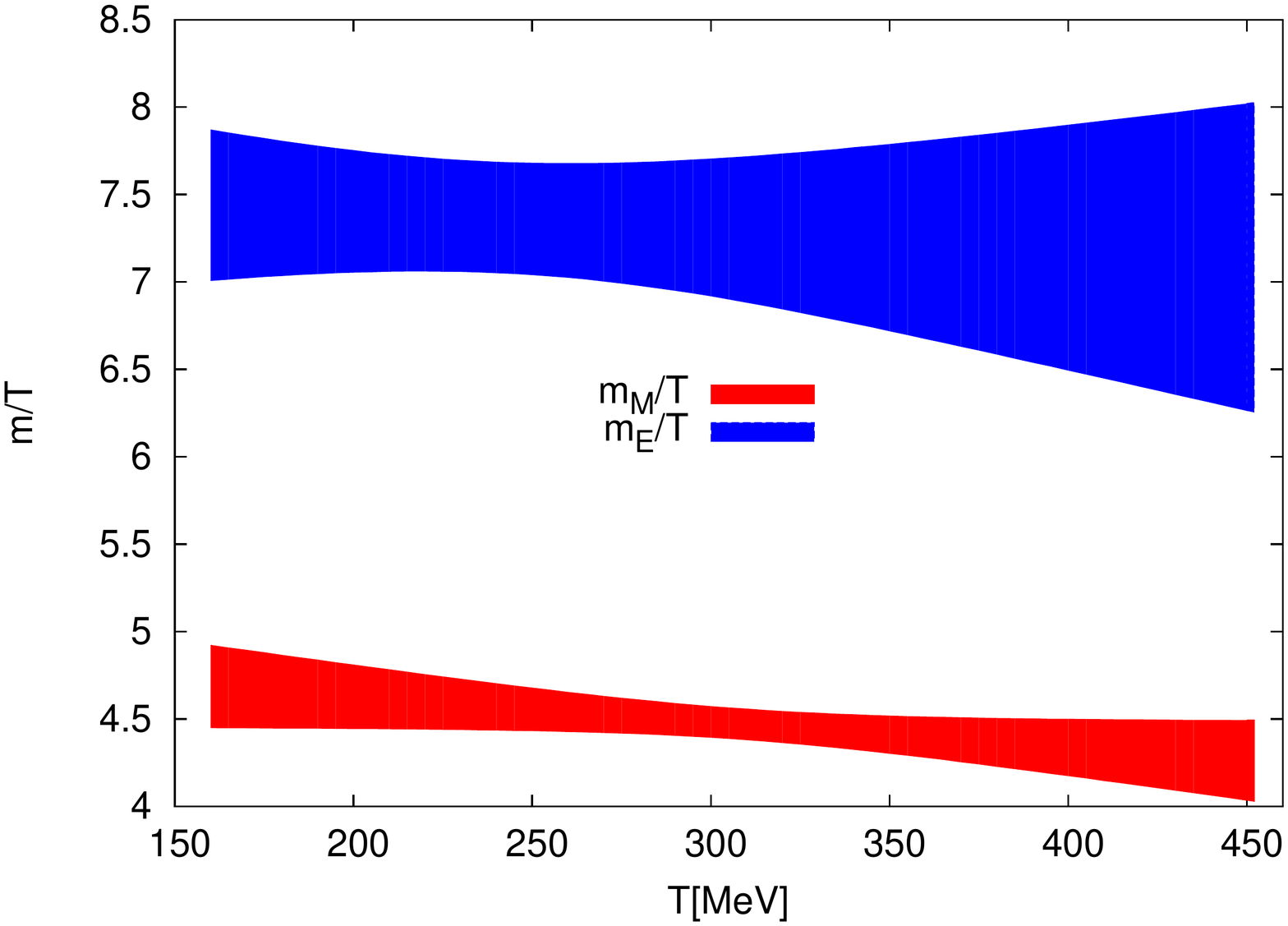}
\includegraphics[angle=0, width=0.48\linewidth]{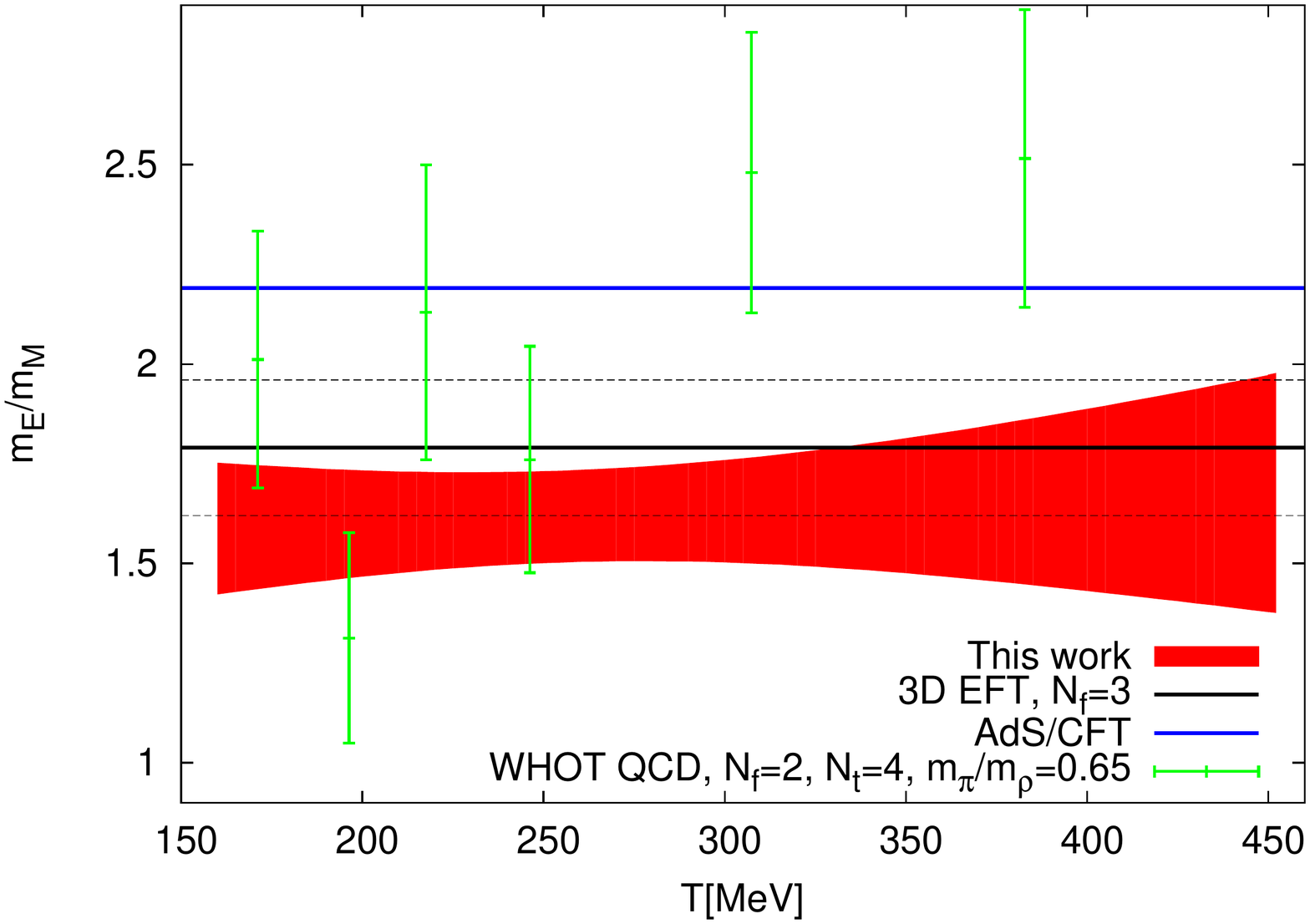}
\caption{The continuum extrapolations of the screening masses and the ratio of the screening masses.
For the ratio $m_E/m_M$ we also included different estimates from the literature:
Lattice results from Ref.~\cite{Maezawa:2010vj}, dimensionally reduced 3D effective
field theory results at $T=2T_c$ from Ref.~\cite{Hart:2000ha}, and results from
$\mathcal{N}=4$ SYM plasma with AdS/CFT from Ref.~\cite{Bak:2007fk}.
}
\label{fig:mass_contlim}
\end{figure}

We continue with the discussion of the electric and magnetic screening masses obtained from the correlators 
(\ref{eq:Cmag}) and (\ref{eq:Cel}). For this analysis we only use lattices above the (pseudo)critical
temperature, since that is the physically interesting range for screening. Next, we mention that for this analysis,
we only use the data with HYP smearing, since we are especially interested in the large r behavior.
To chose the correct fit interval for the Yukawa ansatz, we use the Kolmogorov-Smirnov test to 
check whether the $\chi^2$-s are properly distributed. Note that the determination of the screening
masses does not need additional renormalization. We fit linear functions to all screening masses at all values of $N_t$,
and use these to do a continuum extrapolation from the $N_t=8,10,12$ lattices. The final results, along with
some comparisons with results from the literature are in Fig. \ref{fig:mass_contlim}. For details on the
fitting procedure and the systematic error estimation see \cite{Borsanyi:2015yka}.

\subsection{Comparison with the literature}

We finish this section by comparing our results to those from earlier approximations in the 
literature. For comparison let us use our results at $T=300\rm{MeV} \approx 2 T_c$. Here we have:
\begin{itemize} 

\item This work: 2+1 flavour lattice QCD at the physical point after continuum extrapolation:
\begin{eqnarray*}
m_E/T=7.31(25)    \quad m_M/T=4.48(9)  \\
m_E/m_M=1.63(8)
\end{eqnarray*}

\item Ref.~\cite{Maezawa:2010vj}: 2 flavour lattice QCD with Wilson quarks, 
a somewhat heavy pion $m_{\pi}/m_{\rho}=0.65$, no continuum extrapolation
\begin{eqnarray*}
m_E/T=13.0(11)    \quad m_M/T=5.8(2)  \\
m_E/m_M=2.3(3)
\end{eqnarray*}

\item From Table 1 of Ref.~\cite{Bak:2007fk}: $\mathcal{N}=4$ SYM, large $N_c$ limit, AdS/CFT 
\begin{eqnarray*}
m_E/T=16.05    \quad m_M/T=7.34  \\
m_E/m_M=2.19
\end{eqnarray*}

\item From Figure 3 of Ref.~\cite{Hart:2000ha}: dimensionally reduced 3D effective theory, $N_f=2$ massless quarks
\begin{eqnarray*}
m_E/T=7.0(3)   \quad m_M/T=3.9(2)  \\
m_E/m_M=1.79(17)
\end{eqnarray*}

\item From Figure 3 of Ref.~\cite{Hart:2000ha}: dimensionally reduced 3D effective theory, $N_f=3$ massless quarks
\begin{eqnarray*}
m_E/T=7.9(4)    \quad m_M/T=4.5(2)  \\
m_E/m_M=1.76(17)
\end{eqnarray*}

\end{itemize}

We note, that our results are closest to the results from dimensionally reduced effective
field theory.

\section{Summary}

In this paper we have determined the renormalized static quark-antiquark free energies 
in the continuum limit. We introduced a two step renormalization procedure using only the finite 
temperature results. The low radius part of the free energies tended to the same curve,
corresponding to the expectation that at small distances, the physics is temperature 
independent. We also calculated the magnetic and electric screening masses, from the real 
and imaginary parts of the Polyakov loop respectively. As expected, both of
these masses approximately scale with the temperature as $m \propto T$, with $m_M<m_E$, therefore, magnetic 
contributions dominating at high distances. The values we got for the screening masses are close to 
the values from dimensionally reduced effective field theory.

\section*{Acknowledgment}
Computations were carried out on GPU clusters at the
Universities of Wuppertal and Budapest as well as on supercomputers in
Forschungszentrum Juelich. \\

This work was supported by the DFG Grant SFB/TRR 55, ERC no. 208740. and 
the Lendulet program of HAS (LP2012-44/2012).\\

\end{document}